\documentclass[twocolumn,twoside,slac_two]{revtex4}
\usepackage{graphicx}
\usepackage{fancyhdr}
\usepackage{verbatim}
\input{babarsym}
\def\Ntautau                 {\ensuremath{N_{\scriptscriptstyle{\tau\tau}}}\xspace}

\def\SigEff                  {\ensuremath{\varepsilon_{\scriptscriptstyle{\rm sig}}}\xspace}

\def\Kstar                   {\ensuremath{K^{*}}\xspace}
\def\Kstarp                  {\ensuremath{K^{*+}}\xspace}

\def\dz                      {\ensuremath{d_{\scriptscriptstyle 0}\xspace}}

\def\ie                      {\mbox{i.e.}}

\def\etag                    {$e$-tag}
\def\mutag                   {$\mu$-tag}

\def\mc                      {Monte Carlo}

\def\crosssection            {cross-section}

\usepackage{subfigure}
\usepackage{epsfig}
\usepackage{amsmath}
\usepackage{amssymb}
\usepackage{verbatim}
\usepackage{rotating}
\usepackage[running]{lineno}
\begin{document}

\title{Selected topics in tau physics from \babar}

%

\author{S.Paramesvaran (on behalf of the \babar\ Collaboration)}
\affiliation{Department of Physics, Royal Holloway, University of London, 
Egham, Surrey, TW20 0EX, UK}

\begin{abstract}

Selected results from $\tau$ analyses performed using the \babar\ detector
at the SLAC National Accelerator Laboratory are presented. A precise
measurement of the $\tau$ mass and the $\tau^{+}  \tau^{-}$ mass
difference is undertaken using the hadronic decay mode 
$\tau^{\pm} \rightarrow \pi^{+}\pi^{-}\pi^{\pm}\nu_{\tau}$. In addition
an investigation into the strange decay modes 
$\tau^{-}\rightarrow K^{0}_{S}\pi^{-}\pi^{0}\nu_{\tau}$ and 
$\tau^{-}\rightarrow K^{0}_{S}\pi^{-}\nu_{\tau}$ is also presented, including
a fit to the $\tau^{-}\rightarrow K^{0}_{S}\pi^{-}\nu_{\tau}$ invariant
mass spectrum. Precise values for M(K*(892)) and $\Gamma$(K*(892)) are 
obtained.

\end{abstract}

\maketitle

\thispagestyle{fancy}

\section{Introduction}

\label{sec:intro}
Although the \babar\ detector \cite{Aubert:2002XX} was conceived 
as an experiment to test 
CP-violation in the B meson system, the cross-section for $\tau$ pairs
(0.9\,nb) is 
almost as high as that for $\b\bar{b}$(1.1\,nb). 
This makes \babar\ an excellent place to study $\tau$ physics.  
I present preliminary results from studies of tau decays 
to hadronic final 
states which provide precise measurements on the $\tau$ mass, 
the $\tau^{+}\tau^{-}$ mass difference (Section 2), 
and information on the branching
ratios and mass spectra of the strange
hadronic decays $\tau^{-}\rightarrow K^{0}_{S}\pi^{-}\nu_{\tau}$
and $\tau^{-}\rightarrow K^{0}_{S}\pi^{-}\pi^{0}\nu_{\tau}$ (Section 3).

\section{Precise measurement of $\tau$ mass and $\tau^{+}  \tau^{-}$ mass
difference}

A key test of CPT invariance is to measure the difference in mass between
a particle and its antiparticle. Using 423 fb$^{-1}$ of data from the 
\babar\ detector, a pseudomass endpoint method was used to measure the mass
of the $\tau$ lepton \cite{Aubert:2009ra}. The significant advantage in using this method is that 
it allows the mass of the $\tau^{+}$ and $\tau{-}$ to be measured 
independently, which allows us to test CPT invariance.

The current world average of  $\tau$ mass is 1776.84$\pm$ 0.17 MeV$/c^2$ 
\cite{pdg:2008}, and the
mass difference is $M_{\tau^{+}} - M_{\tau^{-}} < 2.8 \times 10^{-4}$ at 90$\%$
confidence level.

The pseudomass endpoint method was first used by the ARGUS \cite{argus}
collaboration,
and has since been employed by BELLE \cite{belle07}. The premise is first to 
consider reconstructing the mass of the $\tau$ from the final state
hadronic products ($\tau^{-} \rightarrow h^{-}\nu_{\tau}$):

\begin{equation}
M_{\tau} = \sqrt{M_{h}^{2} + 2(\sqrt{s}/2 - E_{h}^{*})(E_{h}^{*} - P_{h}^{*}\cos\theta^{*})} ,
\end{equation}

\noindent where $M_{h}$, $E_{h}$ and $P_h$ are the mass, energy, and magnitude
of the three-momentum of the hadronic system $h$, respectively, $\theta^*$ is
the angle between the hadronic system and the $\nu_\tau$; the * represents
quantities in the $e^{+}e^{-}$ center-of-mass frame. The relationship 
$E_{\tau}^{*} = \sqrt{s}/2$ is used, where $\sqrt{s}$ is the initial $e^{+}e^{-}$
CM energy (10.58 GeV), and $E_{\tau}^{*}$ is the energy of the $\tau$.
In the above representation the angle $\theta^*$ is unknown as the neutrino
escapes undetected; we therefore define the pseudomass $M_{\rm p}$ with the condition
that $\theta^{*} = 0$. This simplifies the above equation:

\begin{equation}
M_{p} =\sqrt{M_{h}^{2} + 2(\sqrt{s}/2 - E_{h}^{*})(E_{h}^{*} - P_{h}^{*})} .
\end{equation}
 
\begin{figure}[h]
\includegraphics[width=80mm]{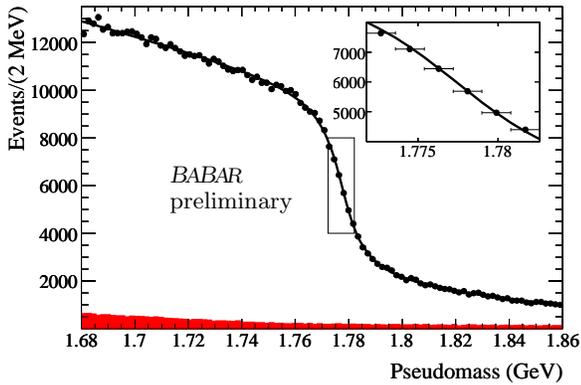}
\put(-160.0,80.0){\babar} 
\put(-160.0,70.0){preliminary}
\caption{Combined $\tau^{+}$ and $\tau^{-}$ pseudomass endpoint distribution.
The points show the data, the curve is the fit to
the data, and the solid area is the background. The inset is
an enlargement of the boxed region around the edge position
showing the fit quality where p1 is most sensitive.}
\label{fig:pseudo}
\end{figure}

\noindent The distribution of $M_p$ has a sharp kinematic cutoff at $M_p = M_\tau$, although
there will be smearing due to initial and final state radiation, and limited
detector resolution.
The signal mode chosen for this study is $\tau^{\pm} \rightarrow \pi^{+}\pi^{-}
\pi^{\pm}\nu_{\tau}$, due in part to its high branching fraction,
(9.03 $\pm$ 0.06) $\times 10^{-2}$ \cite{pdg:2008}, and the
ability to obtain high signal purity. The data are then fitted to an 
empirical function:

\begin{equation}
F(x) = (p_{3} + p_{4}x)\tan^{-1}\frac{\left(p_{1} - x\right)}{p_{2}} + p_{5} + p_{6}x , 
\end{equation}

\noindent where the $p_{i}$ are the parameters and $x$ is the pseudomass.
 A relationship between $p_{1}$ ,the endpoint parameter, and the
pseudomass is required. This is determined from Monte Carlo studies; three
different samples are generated with different tau mass values and these
are fitted with the function defined above. A straight line is then fitted
to the resulting $p_1$ values, which provides the relation to $M_\tau$. The 
data are split into two samples according to the total charge of the 3$\pi$
hadronic final state, and each sample is analyzed independently. The results
of these fits are shown in Figure \ref{fig:pseudo}. This yields an $M_\tau$ of 1776.68 
$\pm$ 0.12(stat)\,MeV$/c^2$.
A number of systematic effects are investigated, including potential
uncertainties on energy loss measurements for charged particles, and 
uncertainties in the magnetic field. However, the dominant uncertainty 
is found to be due to an underestimation of the reconstructed track momenta
in the detector model. This contributes 0.39 MeV to a total systematic
uncertainty of 0.41 MeV.

\section{Strange hadronic tau decays}

Strange hadronic tau decays offer a very clean environment for studying
the weak current. The branching ratios feed directly into a measurement
of $V_{us}$, and fits to the mass spectra can yield resonance parameter
values which can further our understanding of the dynamics of these
systems. 
In this section studies are presented of the hadronic mass distributions 
for the decays $\tau^{-}\rightarrow K^{0}_{S}\pi^{-}\nu_{\tau}$ and
$\tau^{-}\rightarrow K^{0}_{S}\pi^{-}\pi^{0}\nu_{\tau}$ (throughout
the note, charge conjugate modes are implied). A fit to the invariant mass
spectrum from $\tau^{-}\rightarrow K^{0}_{S}\pi^{-}\nu_{\tau}$ is presented
along with precise resonance parameter values of the dominant K*(892)$^-$.
Due to this mass spectrum having a peaking background from 
$\tau^{-}\rightarrow K^{0}_{S}\pi^{-}\pi^{0}\nu_{\tau}$, the hadronic mass
spectra and branching ratio from this mode were also measured and the results
used directly to improve our Monte Carlo modelling. 

\subsection{Analysis of $\tau^{-}\rightarrow K^{0}_{S}\pi^{-}\pi^{0}\nu_{\tau}$
}
To select events of the type $e^+e^- \rightarrow \tau^+\tau^-$ with
one tau-lepton decaying to $K^{0}_{S}\pi^{-}\pi^{0}\nu_{\tau}$, the event is
first divided into two hemispheres in the center-of-momentum system
(CMS) using the thrust axis.  One hemisphere of the event is required
to contain only one charged track; this is defined as the tag
hemisphere.  The other hemisphere is required to have three charged
tracks; this is called the signal hemisphere.  The tag track and at
least one of the signal hemisphere tracks are required to originate from 
the interaction point.

Approximately 35\% of $\tau$-leptons decay to fully leptonic final
states.  Requiring the track in the tag hemisphere to be identified as
an electron or muon while requiring the signal hemisphere to contain
only hadrons strongly reduces backgrounds from $e^+e^- \rightarrow
q\overline{q}$ events.  
Electrons are identified using specialized likelihood selectors, 
whereas a neural network is used to identify muon tracks.

$K^0_{S}$ candidates are constructed from any two oppositely charged
tracks with an invariant mass within $25 \, \mbox{MeV}/c^2$ of the
$K^0_S$ mass, $497.672 \,
\mbox{MeV}/c^2$ \cite{Yao:2006px}.  Only events with exactly one
$K^0_S$ candidate are retained.  The track from the signal side not
originating from the $K^0_S$ candidate is required to be identified as
a pion and originate from the interaction point.  Pions are identified
by $dE/dx$ in the tracking system, the shape of the shower in the
calorimeter and information from the DIRC.  All tracks on the signal
side are required to lie within the geometrical acceptance region of
the EMC and DIRC to ensure good particle identification.

In addition, the net charge of the event must be zero and the thrust
of the event is required to be greater than 0.85 to reduce
the non-$\tau$ background.

\begin{figure}[htbp]
\begin{center}
\includegraphics[width=80mm]{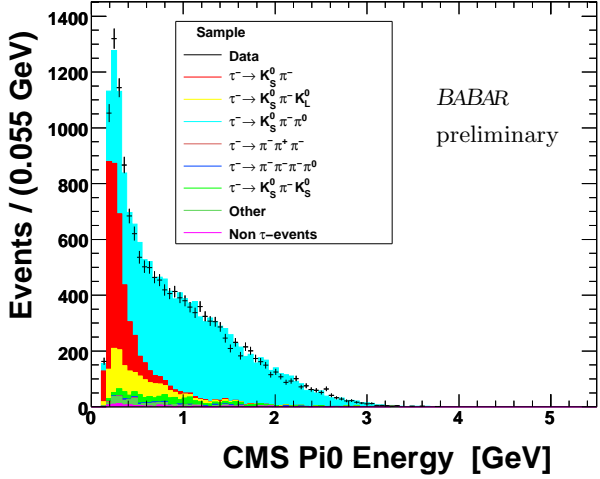}
\put(-63.0,146.0){\babar}
\put(-63.0,132.0){preliminary}
\caption{Distribution of the $\pi^0$ energy.}
\label{fig:pi0energy}
\end{center} 
\end{figure}

Backgrounds from Bhabha events are suppressed by requiring the
momentum of the tag-side track to be less than $4.9 \,
\mbox{GeV}/c$.  Backgrounds from radiative Bhabha and $\mu$-pair
events with a converted photon are suppressed by requiring the modulus
of the cosine of the decay angle to be less than 0.97.  The decay
angle is defined as the angle between the momentum of the $\pi^+$
originating from the $K^0_S$ in the $K^0_S$'s rest frame and the
$K^0_S$ momentum vector in the laboratory frame.  When this quantity
is calculated for $e^+e^-$ conversion pairs misidentified as pions,
its value is concentrated near $\pm 1$.  From studies of missing
transverse event energy, backgrounds from two-photon events are
determined to be negligible.
We also require exactly one identified $\pi^0$ in the event; the trajectory 
of the $\pi^0$ must be within 90 degrees of the $ K^{0}_{S}\pi^{-}$ momentum 
vector. This ensures that the $\pi^0$ is more likely to be from the same 
$\tau$ as the $K_S\pi$.
The neutral energy not attributed to the $K^0_S$ or 
the $\pi^0$ must be less then 100 MeV.  This should be very small 
anyway, but the cut is to reject unwanted photons.
The energy of the $\pi^0$ in the center-of-mass system 
must be greater than 1.2 GeV. This cut is to remove the
large background contribution in the region below 1.2 GeV.
\noindent Figure.~\ref{fig:pi0energy} shows the distribution 
of the $\pi^0$ energy.  

\noindent The branching fraction $\mathcal{B}(\tau^{-}\rightarrow
\bar{K^{0}}\pi^{-}\pi^{0}\nu_{\tau})$ is estimated by

\begin{equation}
\mathcal{B}(\tau^{-}\rightarrow \bar{K^{0}}\pi^{-}\pi^{0}\nu_{\tau}) = 
\frac{1}{2\Ntautau} \frac{N_{\scriptscriptstyle{\rm data}}
               - N_{\scriptscriptstyle{\rm bkg}}}{\SigEff^{'}},
\label{eq:BR}
\end{equation}

\noindent where $\Ntautau$ is the total number of \tautau\ pairs in the data, 
$N_{\scriptscriptstyle{\rm data}}$ is the number of selected
events in data, $N_{\scriptscriptstyle{\rm bkg}}$ is the number
of background events estimated from \mc, and $\SigEff^{'}$ is the corrected
signal efficiency to include $K_{S}^{0}$ and $K_{L}^{0}$ mesons.  

\begin{table}[htbp]
\begin{center}
\caption{\footnotesize{$\mathcal{B}(\tau^{-}\rightarrow \bar{K^{0}}\pi^{-}\pi^{0}\nu_{\tau})$ 
measured in this analysis.}}
\label{tab:kpi0BR}
\begin{tabular}{@{}ll}
\hline
   Sample     &    $\mathcal{B}(\tau^{-}\rightarrow \bar{K^{0}}\pi^{-}\pi^{0}\nu_{\tau})$ [\%] \\
\hline
\etag         & $ 0.353 \pm 0.008 \, \stat \pm 0.016 \, \syst$ \\
\mutag        & $ 0.329 \pm 0.008 \, \stat \pm 0.016 \, \syst$ \\
Combined      & $ 0.342 \pm 0.006 \, \stat \pm 0.015 \, \syst$ \\
\hline
\end{tabular}
\end{center}
\end{table}

The hadronic mass distributions for the different combinations of final
state hadrons are also extracted, and used to tune our \mc. Figure
\ref{fig:newmc} below shows the mass distribution for $K_{S}^{0}\pi^{-}$ from 
$\tau^{-}\rightarrow K^{0}_{S}\pi^{-}\pi^{0}\nu_{\tau}$, overlayed with the 
new Monte Carlo generated using this analysis. 

\begin{figure}[htbp]
\begin{center}
\includegraphics[width=80mm]{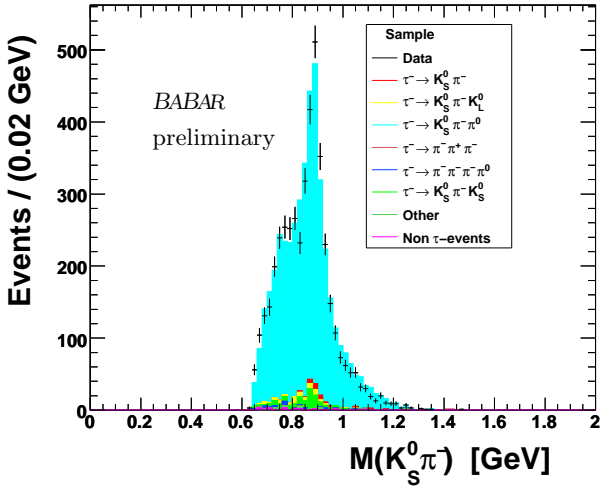} 
\label{lab:newmasskspibg}
\put(-173.0,146.0){\babar}
\put(-173.0,132.0){preliminary}
\caption{The data (points) and tuned Monte Carlo (blue) prediction for the 
$K_{S}^{0}\pi$ mass distribution from the 
$\tau^{-}\rightarrow K^{0}_{S}\pi^{-}\pi^{0}\nu_{\tau}$ signal mode.}
\label{fig:newmc}
\end{center}
\end{figure}

\subsection{Fit to $\tau^{-}\rightarrow K^{0}_{S}\pi^{-}\nu_{\tau}$ mass 
spectrum}

The analysis of the decay $\tau^- \rightarrow K^{0}_{S} \pi^-
\nu_{\tau}$ 
is a fit of the hadronic mass distribution to a parametric function
describing the resonant structure.  From this we obtain precise values
for the mass and width of the K*(892) as well as information on other
resonances present in the spectrum.

We denote the number of events found in bin $i$ (without background
subtraction) by $n_i$.  The prediction for the expectation value of $n_i$,
$\nu_i = E[n_i]$, can be written
\begin{equation}
\label{eq:nui}
\nu_i = \sum_{j=1}^M R_{ij} \mu_{j} + \beta_{i} \;,
\end{equation}
\noindent where $\beta_i$ is the expected number of background events, 
$\mu_{j}$ is 
the predicted number of signal events in bin $j$ before detector
effects (the ``true'' distribution), and $R_{ij}$ is a response matrix
that reflects the limited efficiency and resolution of the detector.
The value of $R_{ij}$ is 
\begin{equation}
\label{eq:Rij}
R_{ij} = P(\mbox{found in bin } i \, | \mbox{true value in
bin } \, j) \;,
\end{equation}
\noindent and thus the efficiency for bin $j$ is found by summing over
all bins where the event could be found. 
The predicted number of events in bin $j$ of the true distribution can
be written
\begin{equation}
\label{eq:muj}
\mu_j = \mu_{\rm tot} \int_{{\rm bin}\, j} f(m; \vec{\theta}) \, dm \;,
\end{equation}
\noindent where $m$ denotes the $K^0_{S}\pi^-$ invariant mass
and $\vec{\theta}$ represents a set of parameters.  

The probability density function (pdf) $f(m; \vec{\theta})$ can be
written \cite{Epifanov:2007rf}:
\begin{eqnarray}
f(m; \vec{\theta}) &\propto& \frac{1}{s}{\left(1-\frac{s}{{m_\tau}^2}\right)} 
\left(1+2\frac{s}{{m_\tau}^2}\right) \nonumber \\
&\times& P\left(P^2{|F_V|}^2 + 
\frac{3({m_K}^2 - {m_\pi}^2)^2}{4s(1+2\frac{s}{{m_\tau}^2})}|{F_S|}^2\right) 
\nonumber
\,\,\,\,\,\,\,
\end{eqnarray}
\noindent where $s = m^2$.  Here the vector form factor $F_V$ is given by
\begin{equation}
\label{eq:fv}
F_V= \frac {1}{1+\beta+\gamma}
[BW_{K^1}(s)+\beta BW_{K^2}(s)+\gamma BW_{K^3}(s)] \;. \nonumber
\end{equation}
\noindent This form allows for the K*(892) and two additional vector resonances. 
The quantities $\beta$ and $\gamma$ are complex interference terms
between the resonances, and the {\it BW} terms refer to the to relativistic
Breit-Wigner functions for the specific resonance, given by
\begin{equation}
\label{eq:bwrs}
BW_R(s) = \frac{M_R^2}{s-{M_R^2} + i\sqrt{s}\Gamma_{R}(s)} \;.
\end{equation} 
\noindent The energy dependent width is given by
\begin{equation}
\label{eq:gammars}
\Gamma_{R}(s)= \Gamma_0R\frac{M_R^2}{s}\left(\frac{P(s)}{P(M_R^2)}\right)^{2\ell+1} \;,
\end{equation}
\noindent where 
\begin{equation}
\label{eq:pofs}
P(s) = \frac{1}{2\sqrt{s}}\sqrt{(s-M_+^2)(s-M_-^2)} \;,
\end{equation}
\noindent and where $M_- = M_K - M_\pi$, $M_+ =M_K + M_\pi$, and $\ell$ is 
orbital angular momentum.  Thus one has $\ell = 1$ 
if the $K\pi$ system is from a P-wave (vector), or $\ell = 0$ if the
$K\pi$ system is from an S-wave (scalar).

The scalar form factor requires a different parametric function and can
include contributions from the K$_{0}^{*}$(800) and K$_{0}^{*}$(1430) signals. 
This is
\begin{eqnarray}
\label{eq:fs}
F_S &=& \varkappa\frac{s}{M^2_{K_{0}^{*}(800)}}BW_{K_{0}^{*}(800)}(s) 
\nonumber \\
&+& \lambda\frac{s}{M^2_{K_{0}^{*}(1430)}}BW_{K_{0}^{*}(1430)}(s) \;.
\end{eqnarray}
Each of the background modes is subtracted from the data, and then a
least-squares fit is performed to the resulting mass spectrum. 
\noindent The fit model 
includes $\tau_j$ as a scale factor that relates the luminosity 
of the Monte Carlo sample for mode $j$ to that of the data, and $r_j$ as a
factor that allows for the uncertainty in the prediction of the rate
of the background process.  The best estimate of $r_j$ is equal to
unity, but this is treated as a Gaussian distributed quantity with a
standard deviation equal to the relative uncertainty on the production
rate for the $j$th background mode.

The uncertainties in the values of other nominally fixed model
parameters, e.g., the resonance parameters of the K*(1410), can be
incorporated into the fit in a similar way.  For a given parameter
$\eta$ one has a previously estimated value $\hat{\eta}$ and standard
deviation $\sigma_{\eta}$, taken from the PDG.  One includes in
the minimisation function a Gaussian term in $\eta$ centered about
$\hat{\eta}$ with a standard deviation $\sigma_{\eta}$, and regards
$\eta$ as an adjustable parameter.
 
We also include terms in the minimisation which account for the
uncertainty in the shapes of background mass distributions. 
This is particularly true for the $K^0_S \pi^- K^0_L$ mode, as
it makes a larger contribution and the information on its shape is
based largely on lower-statistics measurements from LEP \cite{aleph:1998}. 
We introduce 
two additional adjustable parameters, 
$\vec{\alpha} = (\alpha_1, \alpha_2)$, which have the
effect of shifting and stretching the shape of the distribution 
\cite{Cowan:AltHist}. This
transformation is applied to the $m_{ij}$ values for the $K^0_S
\pi^- K^0_L$ background mode and the altered values are then used in
the minimisation.
The fitting procedures described above have been carried out using
a variety of hypotheses.
\begin{figure}[htbp]
\begin{center}
\subfigure[]
{\label{lab:onebwfit}\includegraphics[width=80mm]{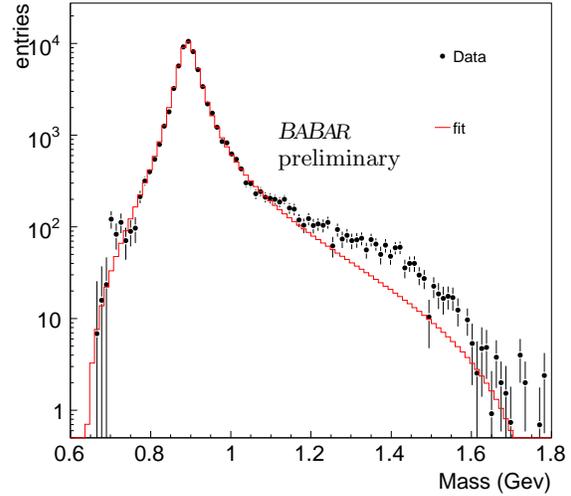}}
\put(-115.0,150.0){\babar} 
\put(-115.0,140.0){preliminary}\\
\subfigure[]
{\label{lab:nominal}\includegraphics[width=80mm]{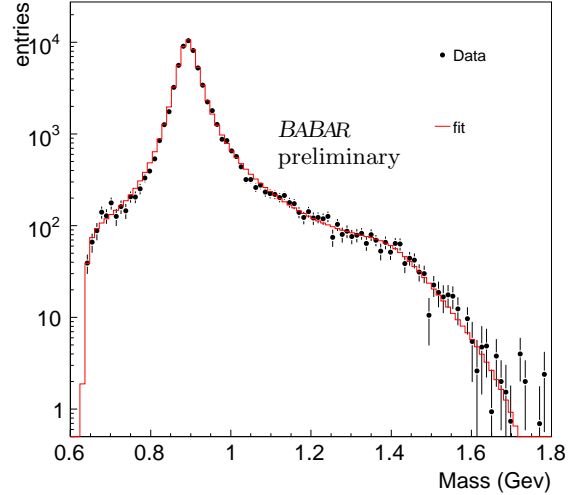}}
\put(-115.0,150.0){\babar} 
\put(-115.0,140.0){preliminary}
\caption{A fit to the $\tau^{-}\rightarrow K^{0}_{S}\pi^{-}\nu_{\tau}$
mass distribution using (a) single K*(892) resonance, (b) a combination of 
K$_{0}^{*}$(800) + K*(892) + K*(1410).}
\end{center}
\end{figure}
Figure~\ref{lab:onebwfit} shows that a single K*(892) is clearly
not enough to model the mass spectrum accurately.  This was seen by
the Belle collaboration \cite{Epifanov:2007rf}, which proposed that the
distribution should contain contributions from a K$_{0}^{*}$(800)  scalar and
K*(1410) vector resonances.

In the region around 1.4 GeV in Fig.~\ref{lab:onebwfit}, the data
are significantly higher than the fitted curve.The addition of the 
K*(1410) gives a significant
improvement to the high mass region, yielding a $\chi^2$ of 130.04 for
95 degrees of freedom.  In these fits the rate of the K*(1410) was allowed to
vary within the error given in the PDG. 

The inclusion of the K$_{0}^{*}$(800)  further reduces our $\chi^{2}$ to 
113.05 for 94 degrees of freedom. This is a significantly better
goodness-of-fit value than our K*(892) + K*(1410) fit model.
For the mass and width of the K$_{0}^{*}$(800)  we use the measurements
from the BES collaboration ~\cite{Ablikim:2005ni}, 
$M=841\pm30^{+81}_{-73}$, $\Gamma=618\pm90^{+96}_{-144}$. 
The result is shown in Fig.~\ref{lab:nominal}.

If instead of a K*(1410) one uses a scalar K$_{0}^{*}$(1430), one
finds a comparable $\chi^2$ value of 114.11 for 94 degrees of
freedom. As such the K*(1410) and K$_{0}^{*}$(1430) cannot be differentiated
on their $\chi^2$ value. To study what combination of K*(1410) and
K$_{0}^{*}$(1430) are present, one could exploit the different spins of
the two resonances by carrying out an angular analysis. This is not part
of the current study.
The resulting values for the mass and width of the K*(892) are found to be
\begin{eqnarray}
M(K^*(892)^-) & = & 894.57 \pm 0.19 \, \mbox{(stat.)} \,\mbox{MeV/c$^2$} 
\nonumber\\
\Gamma(K^*(892)^-) & = & 45.89 \pm 0.43 \, \mbox{(stat.)} \nonumber \;
\,\mbox{MeV/c}
\end{eqnarray}
\noindent The statistical errors quoted already cover a number of
systematic uncertainties such as those in the rates and shapes of
backgrounds, which were incorporated by including corresponding
adjustable parameters in the fit.  Several additional sources of
systematic uncertainty are also taken into consideration.
The response matrix $R_{ij}$ is derived from the Monte Carlo
simulation of the detector. As a conservative estimate of the uncertainty of 
the detector response, which is dominated by  modelling of the
tracking and Calorimeter, we have varied the parameters of the response matrix
by up to $\pm10\%$. As a check of the fitting method we have taken a fully reconstructed
Monte Carlo sample of signal events, and fitted them using the signal
model.  As the MC generator models the $\tau^{-}\rightarrow
K^{0}_{S}\pi^{-}\nu_{\tau}$ decay with only the K*(892) resonance, the
fit model also only contained this resonance.
A further uncertainty in the fit model stems from the choice of
resonances.We take the difference in the mass and width values for the
K*(892) between our nominal fit and the alternative
models which also yield comparable $\chi^{2}$ values, as an estimate of the
systematic uncertainty. 
The quadratic sum of all these sources of systematic uncertainty lead to an 
error on M(K*(892)) and $\Gamma$(K*(892)) of 0.19 MeV/c$^2$ and 0.57 MeV/c
respectively.

\section{Summary and Conclusion}

\noindent Measurements of the $\tau$ mass and $\tau^{+} - \tau^{-}$ mass
difference have been carried out yielding results of:
\begin{eqnarray}
M_{\tau} = 1776.68 \pm 0.12\mbox{(stat.)} \pm 0.41\mbox{(syst.)}\, \mbox{MeV}/c^2.
\nonumber \\*[0.2 cm] 
\frac{M_{\tau^-} - M_{\tau^+}}{\langle M \rangle} = -3.4 \pm 1.3\mbox{(stat.)} \pm 
0.3\mbox{(syst.)} 
\times 10^{-4} \nonumber.
\end{eqnarray}
\noindent where $\langle M \rangle$ is the average of $M_{\tau^{+}}$ and $M_{\tau^{-}}$. 
The $\tau$ mass result is in good agreement with the world average. We also 
find the mass difference result to be consistent with the results published
by the Belle Collaboration at 1.8$\sigma$.

We have also carried out studies of the decays $\tau^{-}\rightarrow
K^{0}_{S}\pi^{-}\nu_{\tau}$ and $\tau^{-}\rightarrow
K^{0}_{S}\pi^{-}\pi^0\nu_{\tau}$ using $384.6 \invfb$ of data.
We have measured the branching ratio for $\tau^{-}\rightarrow
\bar{K^{0}}\pi^{-}\pi^0\nu_{\tau}$, which is found to
be:
\begin{eqnarray}
&{\cal B}(\tau^{-}\rightarrow \bar{K^{0}}\pi^{-}\pi^{0}\nu_{\tau})  = \nonumber \\ &(0.342 \pm 0.006
\, (\mbox{stat.}) \pm 0.015 \, (\mbox{sys.}))\% \;. \nonumber
\end{eqnarray}
For the $\tau^{-}\rightarrow K^{0}_{S}\pi^{-}\pi^0\nu_{\tau}$ mode we
have measured the mass distributions of different
combinations of final state hadrons: $\pi^{-}\pi^{0}$,
$K_{S}^{0}\pi^{-}$, $K_{S}^{0}\pi^{0}$ and $K_{S}^{0} \pi^{-}
\pi^{0}$.  These were used to make important improvements to the
{\tt TAUOLA} Monte Carlo generator, which allowed for a precise
estimation of the background contribution from this mode in the
analysis of the $\tau^{-}\rightarrow K^{0}_{S}\pi^{-}\nu_{\tau}$
channel.

We have carried out a fit of the hadronic mass distribution for 
$\tau^{-}\rightarrow K^{0}_{S}\pi^{-}\nu_{\tau}$. This yields precise
measurements for the mass and width of the K*(892) resonance:
\begin{eqnarray*}
& M(K^*(892)^-)  = \nonumber \\
& 894.30  \pm 0.19 \, \mbox{(stat.)} \pm 0.19 \, \mbox{(syst.)}
\,\mbox{MeV/c$^2$} \;, \\*[0.2 cm]
& \Gamma(K^*(892)^-)  = \nonumber \\ 
& 45.56  \pm 0.43 \, \mbox{(stat.)} \pm 0.57 \, \mbox{(syst.)}
\,\mbox{MeV/c} \;.
\end{eqnarray*}

\begin{figure}[!h]
\begin{center}
{\label{lab:masscomp1}\includegraphics[width=80mm]{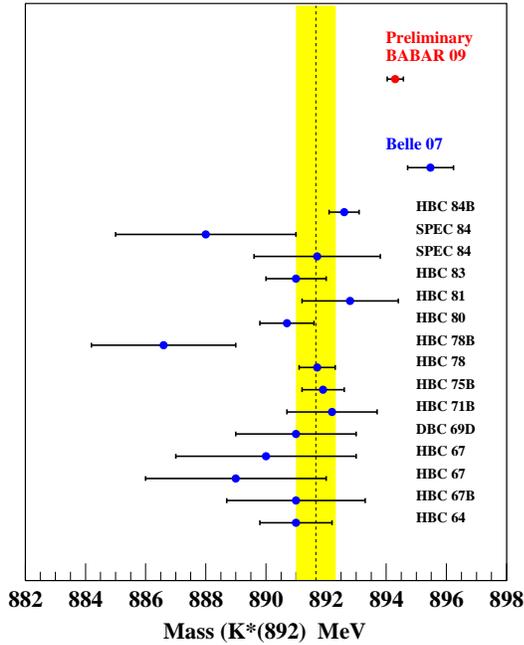}}
{\label{lab:masscomp2}\includegraphics[width=80mm]{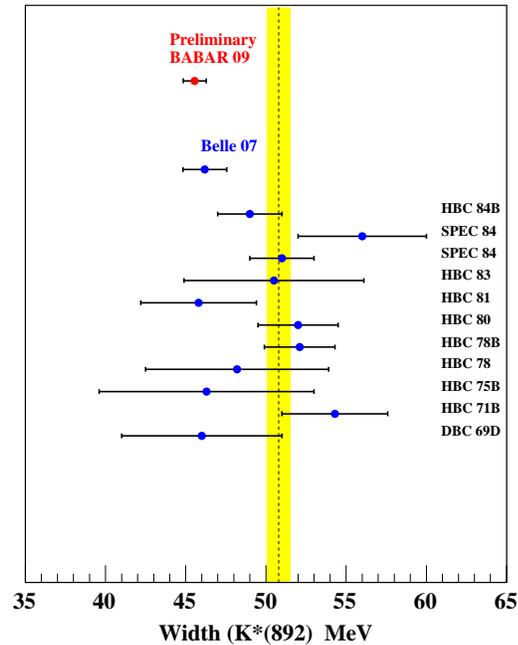}}
\caption{Comparison of the K*(892) mass and width values which were included 
in the PDG~\cite{pdg:2008} calculated average value and the recent result from Belle, 
and our result. The majority of the PDG values are from hydrogen
bubble chamber experiments.}
{\label{lab:comparison}}
\end{center}
\end{figure}

\noindent These values confirm the Belle collaboration's measurements 
\cite{Epifanov:2007rf} that indicated a K*(892) mass several MeV higher 
and a width several MeV lower than the world average.  The results
reported here represent a factor of two improvement in precision
relative to the Belle measurements.
We analyse the possibility of other resonances being present in this
mass spectrum, and conclude that a combination of K*(800), K*(892) and
K*(1410) provides a good description of the data.
Figure \ref{lab:comparison}  shows the results of various measurements that went
into calculating the 2008 PDG average values for the mass and width 
of the K*(892). The Belle 2007 result and our result both indicate a shift
towards 895 MeV for the mass value.


\begin{acknowledgments}
We are grateful for the 
extraordinary contributions of our \pep2\ colleagues in
achieving the excellent luminosity and machine conditions
that have made this work possible.
The success of this project also relies critically on the 
expertise and dedication of the computing organizations that 
support \babar.
The collaborating institutions wish to thank 
SLAC for its support and the kind hospitality extended to them. 
This work is supported by the
US Department of Energy
and National Science Foundation, the
Natural Sciences and Engineering Research Council (Canada),
the Commissariat \`a l'Energie Atomique and
Institut National de Physique Nucl\'eaire et de Physique des Particules
(France), the
Bundesministerium f\"ur Bildung und Forschung and
Deutsche Forschungsgemeinschaft
(Germany), the
Istituto Nazionale di Fisica Nucleare (Italy),
the Foundation for Fundamental Research on Matter (The Netherlands),
the Research Council of Norway, the
Ministry of Education and Science of the Russian Federation, 
Ministerio de Educaci\'on y Ciencia (Spain), and the
Science and Technology Facilities Council (United Kingdom).
Individuals have received support from 
the Marie-Curie IEF program (European Union) and
the A. P. Sloan Foundation.
\end{acknowledgments}

\bigskip 

\end{document}